\documentclass[a4paper,11pt]{article}

\usepackage{graphicx}
\usepackage{dcolumn}
\usepackage{bm}
\usepackage{comment}

\begin{document}

\newcommand{\bin}[2]{\left(\begin{array}{c} \!\!#1\!\! \\  \!\!#2\!\! \end{array}\right)}
\newcommand{\troisjm}[6]{\left(\begin{array}{ccc}#1 & #2 & #3 \\ #4 & #5 & #6 \end{array}\right)}
\newcommand{\sixj}[6]{\left\{\begin{array}{ccc}#1 & #2 & #3 \\ #4 & #5 & #6 \end{array}\right\}}
\newcommand{\neufj}[9]{\left\{\begin{array}{ccc}#1 & #2 & #3 \\ #4 & #5 & #6 \\ #7 & #8 & #9 \end{array}\right\}}

%
%
%

\huge

\begin{center}
Special six-$j$ and nine-$j$ symbols for a single-$j$ shell
\end{center}

\vspace{1cm}

\large

\begin{center}
Jean-Christophe Pain\footnote{jean-christophe.pain@cea.fr}
\end{center}

\normalsize

\begin{center}
CEA, DAM, DIF, F-91297 Arpajon, France
\end{center}


\begin{abstract}
In this Brief Report, we propose general expressions for $6j$ and $9j$ coefficients arising in the study of the $J$-pairing interaction of identical fermions in a single-$j$ shell. In the literature such coefficients are available only for a number of given angular momenta.
\end{abstract}

\section{Introduction}

The present work concerns $6j$ and $9j$ coefficients frequently encountered in nuclear-shell theory. In a series of papers \cite{ZHAO03, ZHAO04, ZHAO05a, ZHAO05b, ZHANG08}, Hamiltonians with attractive interactions between pairs of nucleons coupled to angular momentum $J$ were investigated. The related calculations involve $6j$ and $9j$ coefficients, for which analytical formulas were given only for fixed values of angular momentum $I$ entering as a parameter. In section \ref{sec2}, an explicit form for a $6j$ coefficient used in Ref. \cite{ZHAO04} is presented, together with a sum rule including a weighting factor of the kind $(-1)^J(2J+1)$. In section \ref{sec3}, analytical formulas for two $9j$ coefficients studied in Refs. \cite{ZHAO03, ZHAO05a} are provided.

\section{\label{sec2} six-$j$ coefficients}

In Ref. \cite{ZHAO04}, Zhao and Arima proved that a system of three fermions in a single $j$ shell is exactly solvable in the presence of an angular-momentum $J$-pairing interaction. They found that when the Hamiltonian contains only an interaction between pairs of fermions coupled to spin $J=J_{\mathrm{max}}=2j-1$, the non-integer eigenvalues of three fermions with angular momentum $I$ around the maximum appear as non-integer eigenvalues of four fermions if $I$ is close to or larger than $J_{\mathrm{max}}$. The calculations involve the following $6j$ coefficient:

\begin{equation}\label{eq16j}
\sixj{j}{j}{J}{j}{I}{J},
\end{equation}

for which only expressions for $J=2j-1$ and special cases $I=j-1$, $j$ and $j+1$ are displayed \cite{ZHAO04}. The value of $j$ can be either integer or half-integer. For $J=2j-1$, the $6j$ coefficient presents the particularity that one of the arguments of its first row is smaller by unity than the sum of the two others. Using the following formula given in Ref. \cite {VARSHALOVICH88}:

\begin{eqnarray}
\sixj{a}{b}{a+b-1}{a}{e}{a+b-1}&=&(-1)^{2a+b+e}\nonumber\\
& &2\left\{b(2a+b-1)(2a+b)\right.\nonumber\\
& &\left.-be(e+1)-2a^2\right\}\times\nonumber\\
& &\frac{(2a-1)!(b+e-1)!}{(2a+2b)!(-b+e+1)!},
\end{eqnarray}

with $a=b=j$ and $e=I$, we obtain the following expression

\begin{eqnarray}\label{eq26j}
\sixj{j}{j}{2j-1}{j}{I}{2j-1}&=&(-1)^{3j+I}(9j^2-5j-I(I+1))\times\nonumber\\
& &\frac{(2j)!}{(4j!)}\frac{(j+I-1)!}{(I-j+1)!},\nonumber\\
&=&\frac{(-1)^{3j+I}}{2^{2j}}\frac{(9j^2-5j-I(I+1))}{(4j-1)!!}\times\nonumber\\
& &\frac{(j+I-1)!}{(I-j+1)!},\nonumber\\
& &
\end{eqnarray}

with $n!=n(n-1)(n-2)\cdots$ and $n!!=n(n-2)(n-4)\cdots$. One has $(-1)!!=1$ (occuring for $j=0$) and, when $n$ is even, $n!!=2^{n/2}(n/2)!$. Equation (\ref{eq26j}) is readily obtained by using the Mathematica$^{\mbox{\scriptsize{\textregistered}}}$ calculator in terms of Gamma functions. It is interesting to remind that the $6j$ coefficients of Eq. (\ref{eq16j}) for $I=j$ follow the sum rule \cite{ZHAO03,ZHAO04,ZAMICK05,GINOCCHIO93,TALMI93}:

\begin{equation}\label{eqz1}
\frac{1}{3}\left(\frac{2j+1}{2}+2\sum_{\mathrm{even}\; J}(2J+1)\sixj{j}{j}{J}{j}{j}{J}\right)=\left[\frac{2j+3}{6}\right],
\end{equation}

where $\left[x\right]$ represents the integer part of $x$ (largest integer not exceeding $x$). Similarly, for $I=j+1$, one has

\begin{eqnarray}\label{eqz2}
\frac{1}{3}\left(\frac{2j-1}{2}-2\sum_{\mathrm{even}\; J}(2J+1)\sixj{j}{j}{J}{j}{j+1}{J}\right)=\left[\frac{j}{3}\right].
\end{eqnarray}

Such sum rules are required in order to determine the number of independent interactions in a given $j$ shell that conserve seniority \cite{TALMI93}. More general sum rules are given by Zhao and Arima in the appendix of Ref. \cite{ZHAO04}. The relations (\ref{eqz1}) and (\ref{eqz2}) are, for a half-integer $j$, particular cases of the identity:

\begin{equation}\label{s1}
\sum_{\mathrm{even}\; J}(2J+1)\sixj{j}{j}{J}{j}{I}{J}=
\left\{\begin{array}{ll}
\frac{3}{2}\left[\frac{2I+3}{6}\right]-\frac{I}{2}-\frac{1}{4} & \mbox{if $I\leq j$}\\
\frac{3}{2}\left[\frac{3j-3-I}{6}\right]+\frac{3}{2}\delta_{I,j}-\frac{1}{2}\left[\frac{3j+1-I}{2}\right] & \mbox{if $I\geq j$}
\end{array}\right.,
\end{equation}

where

\begin{equation}
\delta_{I,j}=\left\{
\begin{array}{ll}
0 \;\;\;\;\;\mathrm{if}\;\;\;\;\; (3j-3-I) & \mbox{mod $6=1$}\\
1 & \mbox{otherwise.}
\end{array}\right.
\end{equation}

The same summation over odd values of $J$ leads to:

\begin{equation}\label{s2}
\begin{array}{ll}
\sum_{\mathrm{odd}\; J}(2J+1)\sixj{j}{j}{J}{j}{I}{J} & =\frac{I}{2}+\frac{1}{4}-\frac{3}{2}\left[\frac{2j+3}{6}\right] \\
 & =\left\{\begin{array}{ll}
-1 & \mbox{if $2j=3k$} \\
0 & \mbox{if $2j=3k+1$}\\
1 & \mbox{if $2j=3k+2$}.
\end{array}\right.\end{array}
\end{equation}
 
All the sum rules given in Ref. \cite{ZHAO04}, including Eqs. (\ref{s1}) and (\ref{s2}), involve the weighting factor $(2J+1)$. We would like to mention that Eq. (\ref{s1}) and Eq. (\ref{s2}) can be combined to obtain the following sum rule with the weighting factor $(-1)^J(2J+1)$:

\begin{equation}
\sum_J(-1)^J(2J+1)\sixj{j}{j}{J}{j}{I}{J}=1-(I-r)+3\left[\frac{I-r}{3}\right],
\end{equation}

with $t=(1+(-1)^{2j})/2$ and $r=3(1-t)/2$. Such a relation was also derived by Vanagas and Batarunas in their paper on the characters of the symmetric group SO(3) \cite{VANAGAS61,KANCEREVICIUS90}.

\section{\label{sec3} nine-$j$ coefficients}

In Ref. \cite{ZHAO03}, Zhao \emph{et al.} showed that an attractive $J$-pairing interaction favors pairs with angular momentum $J$ in low-lying states and discovered that a large array of eigenvalues of four nucleons in a single-$j$ shell are asymptotic integers when $J\approx J_{max}=2j-1$ and the total angular momentum $I$ is not very close to $I_{max}=4j-6$. This phenomenon originates from the validity of the pair-truncation scheme and special features of particular $9j$ coefficients, which are of the kind:

\begin{equation}\label{eq1}
\neufj{j}{j}{J}{j}{j}{J}{J}{J}{I},
\end{equation}

the value of $j$ being either integer or half-integer. Such coefficients play a role in the calculation of matrix elements, and in the numbering and classification of states of given spin in the presence of angular-momentum $J$-pairing interaction. It is easy to see that the $9j$ coefficient of Eq. (\ref{eq1}) is equal to zero if $I$ is odd, because a phase factor $(-1)^{4j+4J+I}=(-1)^I$ appears if one exchanges the first and second rows. 

The particular $9j$ coefficient used in appendix A of Ref. \cite{ZHAO03} and corresponding to $J=2j-1$, is equal to

\begin{equation}\label{eq2}
\neufj{j}{j}{2j-1}{j}{j}{2j-1}{2j-1}{2j-1}{I}.
\end{equation}

Zhao \emph{et al.} provide the expression of this coefficient for a few particular cases ($I=0,2,4$ and $6$) in appendix A of Ref. \cite{ZHAO03}, but mention that they were unable to get a universal formula. The latter $9j$ coefficient is also studied in appendix A of Ref. \cite{ZHAO05a}, and formulas are given for particular cases $I=0, 2, 4,\cdots, 12$. The authors explain that they obtained their expressions using the expansion of the $9j$ coefficients of Eq. (\ref{eq1}) in terms of $6j$ coefficients:

\begin{eqnarray}
\neufj{j}{j}{J}{j}{j}{J}{J}{J}{I}&=&\sum_{k=|j-I|}^{j+I}(-1)^{2k}(2k+1)\times\nonumber\\
& &\sixj{j}{j}{J}{J}{I}{k}^2\sixj{j}{j}{J}{j}{k}{J},
\end{eqnarray}

together with analytical expressions for the $6j$ coefficients in the cases where $J=2j-1$ and $2j$ for different fixed values of $I$. We found empirically that the particular $9j$ coefficient of Eq. (\ref{eq2}) is equal to

\begin{eqnarray}\label{eq3}
\neufj{j}{j}{2j-1}{j}{j}{2j-1}{2j-1}{2j-1}{I}&=&\frac{(-1)^{2j+I/2}}{2}\frac{(I-1)!!~(4j-I-3)!!}{(4j+1)!!}\nonumber\\
& &\frac{(2j+I/2-1)!~(2j-1)!}{(I/2)!~(4j-1)!}\nonumber\\
& &\left(8j^2-6j-\frac{I(I+1)}{2}\right),
\end{eqnarray}

where $I$ is an even integer. Equation (\ref{eq3}) is conjectured partly based on Refs. \cite{ZHAO03,ZHAO05a}, and has been checked numerically by using a Fortran code for various $j$ and $I$ up to very large values ($j = 100, I =
156$).

The second $9j$ coefficient described in appendix A of Ref. \cite{ZHAO05a}, corresponding to $J=2j$, has the following expression:

\begin{equation}\label{eq4}
\neufj{j}{j}{2j}{j}{j}{2j}{2j}{2j}{I}.
\end{equation}

The authors give the expression of this $9j$ coefficient for particular cases $I=0, 2, 4,\cdots, 24$. Using the representation of the $9j$ coefficient by the triple sum formula of Ali\v{s}auskas and Jucys \cite{ALISAUSKAS71,WU73,BIEDENHARN81,ZHAO88,RAO89}, Bandzaitis \emph{et al.} \cite{BANDZAITIS64,JUCYS77} found the general formula:

\begin{eqnarray}
\neufj{a_1}{a_2}{a_{12}}{a_3}{a_4}{a_{34}}{a_1+a_3}{a_2+a_4}{b}&=&(-1)^{a_1-a_2-a_3+a_4+b}\times\nonumber\\
& &\left[\frac{(2a_1)!(2a_2)!}{(2a_1+2a_3+1)!(2a_2+2a_4+1)!}\right]^{1/2}\times\nonumber\\
& &\left[\frac{(2a_3)!(2a_4)!}{(a_1+a_2+a_{12}+1)!(a_1+a_2-a_{12})!}\right]^{1/2}\times\nonumber\\
& &\left[\frac{(a_1+a_2+a_3+a_4+b+1)!}{(a_3+a_4+a_{34}+1)!}\right]^{1/2}\times\nonumber\\
& &\left[\frac{(a_1+a_2+a_3+a_4-b)!}{(a_3+a_4-a_{34})!}\right]^{1/2}\times\nonumber\\
& &\troisjm{a_{12}}{a_{34}}{b}{a_1-a_2}{a_3-a_4}{-a_1+a_2-a_3+a_4},\nonumber\\
& &
\end{eqnarray}

where the $9j$ symbol is expressed with a $3jm$ coefficient. In our case, we have $a_1=a_2=a_3=a_4=j$, $a_{12}=a_{34}=2j$ and $b=I$, which leads to:

\begin{eqnarray}
\neufj{j}{j}{2j}{j}{j}{2j}{2j}{2j}{I}&=&(-1)^I\frac{(2j)!^2}{(4j+1)!^2}\times\nonumber\\
& &\sqrt{(4j+I+1)!(4j-I)!}\troisjm{2j}{2j}{I}{0}{0}{0},\nonumber\\
& &
\end{eqnarray}

where the $3jm$ coefficient is \cite{EDMONDS57}

\begin{eqnarray}
\troisjm{2j}{2j}{I}{0}{0}{0}&=&(-1)^{2j+I/2}\frac{I!\left(2j+I/2\right)!}{((I/2)!)^2\left(2j-I/2\right)!}\times\nonumber\\
& &\sqrt{\frac{(4j-I)!}{(4j+I+1)!}}.
\end{eqnarray}

Therefore, the particular $9j$ coefficient of Eq. (\ref{eq4}) is equal to

\begin{eqnarray}\label{eq5}
\neufj{j}{j}{2j}{j}{j}{2j}{2j}{2j}{I}&=&(-1)^{2j+I/2}\left(\frac{(2j)!}{(4j+1)!}\right)^2\frac{I!~(4j-I)!~\left(2j+I/2\right)!}{((I/2)!)^2~\left(2j-I/2\right)!}\nonumber\\
&=&(-1)^{2j+I/2}\frac{(2j)!^2}{(4j)!(4j+1)^2}\frac{\bin{2j}{I/2}\bin{2j+I/2}{I/2}}{\bin{4j}{I}},
\end{eqnarray}

where $I$ is an even integer and $\bin{n}{p}$ represents the binomial coefficient.

\section{Conclusion}

In this paper, we have presented a few formulas of special $6j$ and $9j$ symbols for a single-$j$ shell. Some of these results were given for fixed values of the angular momentum in previous studies \cite{ZHAO03,ZHAO04,ZHAO05a,ZHAO05b,ZHANG08}; in the present work, we proposed unified formulas. 

The results for $6j$ coefficients are proved by a formula given in Ref. \cite{VARSHALOVICH88}. They have the particularity that, on one row, one of the argument is the summation of the two others minus one. They are also readily obtained by using the Mathematica$^{\mbox{\scriptsize{\textregistered}}}$ $6j$ calculator. One of the $9j$ results is proved by using a relation given in Refs. \cite{JUCYS77,BANDZAITIS64} and involving a $3jm$ coefficient; the other is confirmed by computer codes but a mathematical proof is warranted in the future.  

\vspace{0.5cm}

\normalsize {\bf Acknowledgments}

\vspace{0.5cm}

The author would like to thank the anonymous referee for very helpful comments and suggestions.

\end{document}